\documentclass{aa}     

\usepackage{indentfirst}
\usepackage{graphicx}
\usepackage{txfonts}
\usepackage{float}
\usepackage{lipsum}
\usepackage{subcaption}       
\usepackage{hyperref}
\hypersetup{
     colorlinks   = true,
     linkcolor   =  cyan,
     citecolor    = teal,
     urlcolor = blue
}
                                
\usepackage{lscape}            
\usepackage{xcolor}

\usepackage{placeins}          

\makeatletter

\renewcommand*\aa@journalname{}

\renewcommand*\aa@manuscriptname{}

\renewcommand*\aa@textidlineempty{}

\fancypagestyle{otherpage}{%
  \fancyhf{}%
  \fancyhead[CO]{\aa@headfont\aa@headings}%   % odd:  Author: Title  (keep)
  \fancyhead[CE]{}%                            % even: blank
  \fancyfoot[RO]{\aa@footfont\thepage}%        % page number only
  \fancyfoot[LE]{\aa@footfont\thepage}%
  \renewcommand*{\headrulewidth}{\z@}%
  \renewcommand*{\footrulewidth}{\z@}%
}

\makeatother
\DeclareRobustCommand{\Mpc}{\mathrm{Mpc}}
\newcommand{\msun}{\mbox{M$_\odot$}}

\DeclareRobustCommand{\pc}{\mathrm{pc}}
\DeclareRobustCommand{\kpc}{\mathrm{kpc}}

\DeclareRobustCommand{\magarcsec}{\mathrm{mag \ arcsec^{-2}}}

\begin{document}

%%%%%%%%%%%%%%%%%%%%%%%%%%%%%%%%%%%%%%%%
   \title{ Rubin LSST DP2 unveils almost-dark galaxies in the Virgo Cluster }

%%%%%%%%%%%%%%%%%%%%%%%%%%%%%%%%%%%%%%%%

   \author{Minh Ngoc Le \inst{1,2,3}\fnmsep\thanks{Corresponding author: lmngoc1509@gmail.com}
        %\and Ignacio Trujillo \inst{1,2}
        \and Johan H. Knapen \inst{1,2}
        \and Željko Ivezić \inst{4}
        \and Junais \inst{5}
        \and Aaron Watkins \inst{6}
        \and Dan S. Taranu \inst{7}
        \and Pierre-Alain Duc \inst{8}
        \and Anthony Englert \inst{9}
        \and Erfan Nourbakhsh \inst{7}
        \and  J. Anthony Tyson \inst{10}
        \and Jiaxuan Li \inst{11}
        \and Roberto J. Assef \inst{12}
        \and Uzay Aydin \inst{13}
        \and Jose Benavides \inst{14}
        \and Natanael M. Cardoso \inst{15}
        \and Priyanka Chakraborty \inst{16}
        \and Ricardo Demarco \inst{17}
        \and Bililign Dullo \inst{18}
        \and Alister W. Graham \inst{19}
        \and Marc Huertas-Company \inst{1,2,20,21}
        \and Kian-Tat Lim \inst{22}
        \and Jakub Nadolny \inst{23}
        \and Dieu Nguyen \inst{24}
        \and Reynier Peletier \inst{3}
        \and Rossella Ragusa \inst{25}
        \and Kevin Reil \inst{22}
        \and R. Michael Rich \inst{26}
        \and Hector Hernandez-Toledo \inst{27}
        \and Michael H. F. Wilkinson \inst{28}
        }

   \institute{Instituto de Astrofísica de Canarias, Vía Láctea S/N, E-38205 La Laguna, Spain 
   \and Departamento de Astrofísica, Universidad de La Laguna, E-38206 La Laguna, Spain
   \and Kapteyn Astronomical Institute, University of Groningen, P.O. Box 800, 9700AV Groningen, The Netherlands
   \and Department of Astronomy and the DiRAC Institute, University of Washington, 3910 15th Avenue, NE, Seattle, WA 98195, USA
   \and Leibniz-Institut für Astrophysik Potsdam (AIP), An der Sternwarte 16, 14482 Potsdam, Germany
   \and Centre for Astrophysics Research, University of Hertfordshire, College Lane, Hatfield AL10 9AB, UK
   \and Department of Astrophysical Sciences, Princeton University, Princeton, NJ 08544, USA
   \and Université de Strasbourg, CNRS, Observatoire astronomique de Strasbourg, UMR 7550, F-67000 Strasbourg, France
   \and Department of Physics, Brown University, 182 Hope Street, Providence, RI 02912, USA
   \and Physics Department, University of California, One Shields Avenue, Davis, CA 95616, USA
   \and Stanford University, 450 Jane Stanford Way, Stanford, CA 94305, USA
   \and Instituto de Estudios Astrof\'isicos, Facultad de Ingenier\'ia y Ciencias, Universidad Diego Portales, Santiago, Chile
   \and Erciyes University, Department of Astronomy and Space Sciences, Talas, Kayseri, Turkey
   \and Department of Physics and Astronomy, University of California, Riverside, 900 University Avenue, Riverside, CA 92521, USA
   \and Universidade de São Paulo, São Paulo, 05508-010, SP, Brazil
   \and Center for Astrophysics, Harvard \& Smithsonian, 60 Garden Street, Cambridge, MA, 02138, USA
   \and Institute of Astrophysics, Facultad de Ciencias Exactas, Universidad Andr\'es Bello, Sede Concepci\'on, Talcahuano, Chile
   \and Embry-Riddle Aeronautical University, Daytona Beach, FL 32114, USA
   \and Centre for Astrophysics and Supercomputing, Swinburne University of Technology, Hawthorn, VIC 3122, Australia
   \and Université PSL, Observatoire de Paris, Sorbonne Université, CNRS, LERMA, 75014 Paris, France
   \and Université Paris-Cité, 5 Rue Thomas Mann, 75013 Paris, France
   \and SLAC National Accelerator Laboratory, 2575 Sand Hill Rd., Menlo Park, CA 94025, USA
   \and Astronomical Observatory Institute, Faculty of Physics and Astronomy, Adam Mickiewicz University, ul. Słoneczna 36, 60-286 Poznań, Poland
   \and Department of Astronomy, University of Michigan, 1085 South University Avenue, Ann Arbor, MI 48109, USA
   \and INAF – Osservatorio Astronomico di Capodimonte, Salita Moiariello 16, I-80131 Napoli, Italy
   \and Department of Physics and Astronomy, UCLA, Los Angeles, CA, USA
   \and Universidad Nacional Autónoma de México, Instituto de Astronomía, A.P. 70-264, 04510 CDMX, México
   \and Bernoulli Institute of Mathematics, Computer Science and Artificial Intelligence, University of Groningen, Groningen, The Netherlands
   }
   
   \date{\today}
   \authorrunning{Le et al.}
   \titlerunning{Rubin DP2 unveils almost-dark galaxies in Virgo}
   
  \abstract{Galaxies with the faintest surface brightness are currently known only in the Local Group. Similar objects should exist beyond our vicinity and are crucial for understanding galaxy evolution, structure, and dark matter content, yet surveys have not reached the depth required to detect them systematically. We present a population of seven almost-dark galaxies identified in Data Preview 2 of the Rubin Legacy Survey of Space and Time. They surround M49 in the vicinity of the Virgo Cluster, and exhibit central surface brightnesses in the range of $26.9 -28.5 \, \magarcsec{}$ in the $g$-band, with half-light radii of $0.6 - 4.6 \, \kpc{}$ at the distance of Virgo, and stellar masses of $10^6 - 10^7 \, \msun{}$. Their characteristics are analogous to those of the faintest and low-mass galaxies identified among satellite galaxies And~XXI, And~XXIII, and And~XXV in the Local Group. This discovery demonstrates the power of the forthcoming Rubin LSST 10-year survey to uncover extremely faint galaxies at scale, promising the large statistical samples needed to constrain the faint-end luminosity function and the nature of dark matter.}

  % \keywords{Galaxies: clusters: individual: Virgo
  % -- Galaxies: dwarf
   %            }

   \maketitle
   \nolinenumbers

%%%%%%%%%%%%%%%%%%%%%%%%%%%%%%%%%%%%%%%%%%%%%%%%%%%%%%%%%%%%%%
\section{Introduction}

Extreme galaxies such as RCP~32 \citep{2021A&A...656A..44R}, Nube \citep{Nube_2024A&A...681A..15M}, and the recently reported TTT~J1237327+143535 in the Virgo Cluster \citep{TTT_AD} demonstrate that analogues of galaxies in the very low surface brightness, low-mass end of the Local Group satellites \citep{McConnachie_2012} exist beyond the Local Group. Those analogues, at large distances, with extremely low surface brightness ($ \mu_{0,g} \sim 27 \, \magarcsec$ and fainter) and large effective radii (one to a few $\kpc{}$), are invisible to all but the deepest imaging surveys as the Next Generation Virgo Cluster Survey (NGVS; \citealt{2012ApJS..200....4F}), the Large Binocular Telescope Imaging of Galactic Halos and Tidal Structures (LIGHTS; \citealt{Trujillo_LIGHTS_2021} and  \citealt{2024AJ....168...69Z}), Euclid \citep{2025A&A...697A...1E}, and the Rubin 10-year Legacy Survey of Space and Time (Rubin LSST; \citealt{2019ApJ...873..111I}). Following the definition in \citet{Nube_2024A&A...681A..15M}, we refer to this class of objects as almost-dark galaxies (ADs). While ADs occupy a similar effective radii range to ultra-diffuse galaxies (UDGs, \citealt{2015ApJ...798L..45V}), they are typically 10--100 times fainter, placing them in a physically distinct and largely unexplored regime at the extreme faint end of the galaxy luminosity function.

Detecting ADs and precisely quantifying their physical properties are challenging due to their low surface brightness. Additionally, beyond the Local Group, the extensions of ADs make it difficult to distinguish them from background galaxies. Expanding the census of objects like TTT~J1237327+143535 sheds light on the physical processes that suppress star formation in low-mass halos — whether environmental quenching, reionisation feedback, or internal processes — and on how galaxies at the extreme faint end of the luminosity function form and evolve.

% what is lack in literature - new in this paper
In this work, we report the discovery of seven AD candidates in the Virgo Cluster with Rubin LSST Data Preview 2 (DP2). These ADs have central surface brightness in the $g$-band of $27 \, \magarcsec{}$ and fainter, with half-light radii between $0.6 - 4.6\, \kpc{}$. Among them, Rubin~J122703.1+080018 (AD~7), if confirmed, will be one of the most extreme galaxies ever detected with the combination of a flat core-Sérsic profile, large effective radius ($r_\mathrm{e}\sim 4.6\, \kpc{}$), and extremely low surface brightness ($\mu_{0,g} \sim 28.5 \, \magarcsec{}$). This type of object is at the edge of current observational technology. Throughout this work, we used the distance to the Virgo Cluster $\mathrm{D} = 16.17 \, \Mpc{}$ ($\pm 0.25 \, \Mpc{}$ statistical error and $\pm 0.47\, \Mpc{}$ systematic error, \citealt{2025ApJ...982...26A}).

%%%%%%%%%%%%%%%%%%%%%%%%%%%%%%%%%%%%%%%%%%%%%%%%%%%%%%%%%%%%%%
\section{M49 - the "Cosmic Treasure Chest" in Rubin LSST DP2}\label{section:data }

The NSF-DOE Vera C. Rubin Observatory, located on Cerro Pachon in Chile, started its ten-year survey in June 2026. The recent data release DP2 (July 27, 2026) includes commissioning and early operations observations taken between April 2025 and January 2026, and the  Rubin First Look images. DP2 contains data from six filters ($u$, $g$, $r$, $i$, $z$, $y$), with total sky coverage for co-add images of $3000 \, $. All the fields in DP2 were observed using the LSST Camera \citep{10.71929/rubin/2571927} - the same camera used for the LSST 10-year imaging. More details of Rubin DP2 are in \citet{RTN-115}. Details on the Rubin pipeline software can be found at \citet{PSTN-019}.

The "Cosmic Treasure Chest" field is part of Rubin First Look images. This field covers an area of about $24 \, \deg^2$, and is centered at RA $= 186^\circ$, Dec $= 6.5^\circ$ (near M49 - the central galaxy of the Virgo~B subcluster). The field was observed with $u$, $g$, $r$, and $i$ filters, with the total exposure time per target $0.3, 2.0, 3.0,$ and $ 1.7$ hours in $u$, $g$, $r$, and $i$ bands, respectively \citep{pai2026deeplookultrafaintmilky}. The mean surface brightness limit of this field at the search areas reaches $29.5, 30.8, 30.4, 29.7 \, \magarcsec{}$, correspondingly, for $u$, $g$, $r$, and $i$ bands ($3\sigma$ in areas equivalent to boxes of $10''\times 10''$), similar to the expected depth of the Rubin 10-year coadd. The $5\sigma$ point-source limit, calculated within apeatures of $2\times \mathrm{FWHM}$ (with FWHM $\sim 1.3''$), for each filter reaches $25.2 \, \mathrm{mag}$ ($u$ band), $26.4 \, \mathrm{mag}$ ($g$ band), $26.0 \, \mathrm{mag}$ ($r$ band) and $25.4 \, \mathrm{mag}$ ($i$ band) -- about $0.5 - 1 \, \mathrm{mag}$ shallower than the predictions for Rubin 10-year coadd.

\begin{figure*}[ht!]
    \centering
    {\includegraphics[width=1\linewidth] {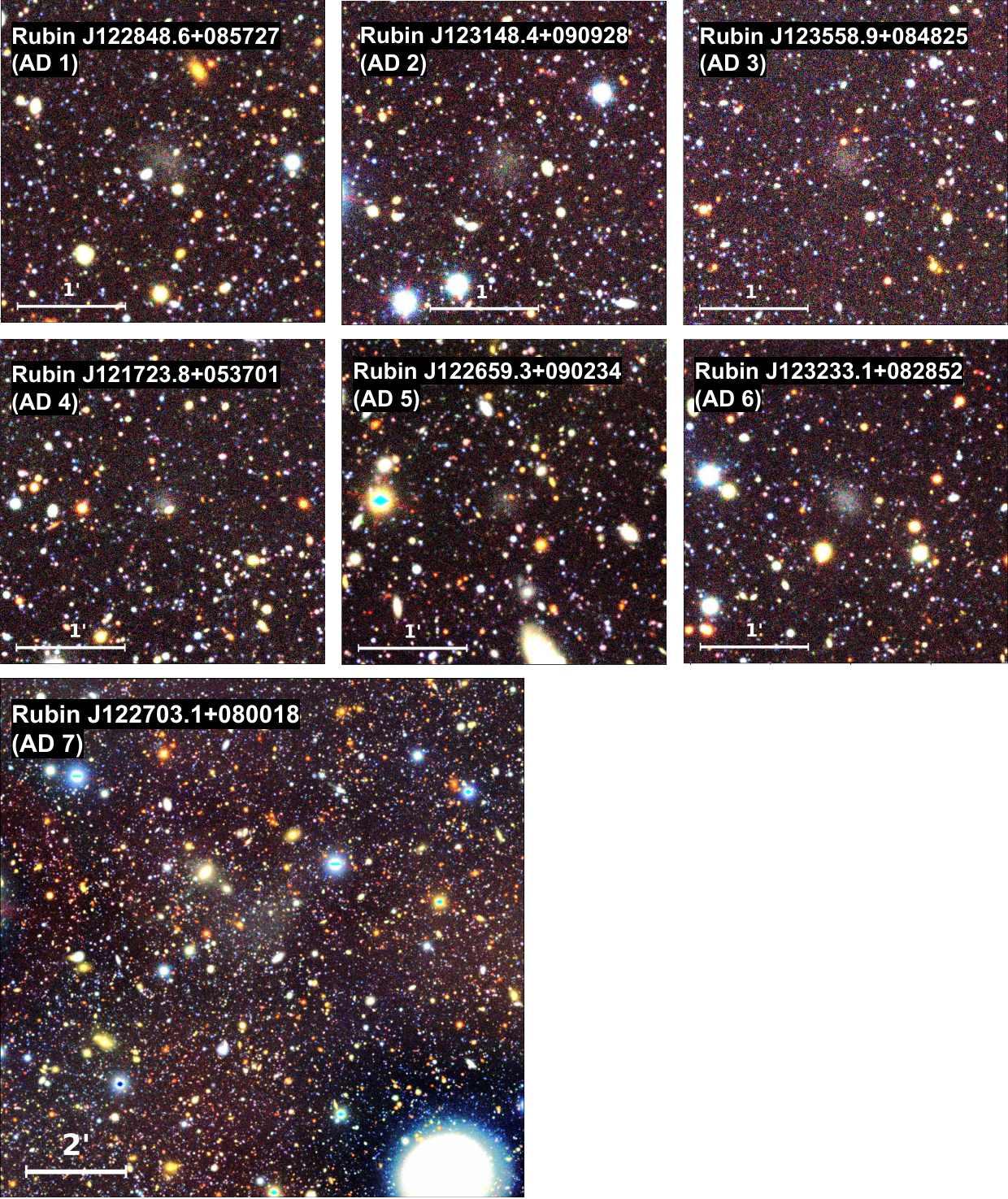}}
      \caption{Colour images combined $g$-, $r$-, and $i$- bands produced with \textit{astscript-color-faint-gray} \citep{2024RNAAS...8...10I} of seven ADs in Rubin DP2. One arcminute corresponds to $4.7 \, \kpc{}$ at the Virgo Cluster distance. North is up and East to the left.}
    \label{fig:dwarfs color image}
    \end{figure*}

We found seven AD candidates during visual inspection around this field, with their color images shown in Fig. \ref{fig:dwarfs color image}: Rubin~J122848.6+085727 (AD~1), Rubin~J123148.4+090928 (AD~2), Rubin~J123558.9+084825 (AD~3), Rubin~J121723.8+053701 (AD~4), Rubin~J122659.3+090234 (AD~5) - which is independently discovered and reported in \citet{mitra2026rubinj1226594090236extremelylow}, Rubin~J123233.1+082852 (AD~6), and Rubin~J122703.1+080018 (AD~7) \footnote{Objects AD~2, AD~3, and AD~6 were independently identified in our analysis prior to the release of the recent NGVS catalog \citep{Ferrarese_2026} in which they also appear.}. The projected distances of these ADs to M49 are within the virial radius of M49 ($r_\mathrm{vir} \sim 1 \, \Mpc{}$); therefore, we assume they are associated with the Virgo Cluster.

\begin{figure*}[ht!]
    \centering
    {\includegraphics[width=1\linewidth] {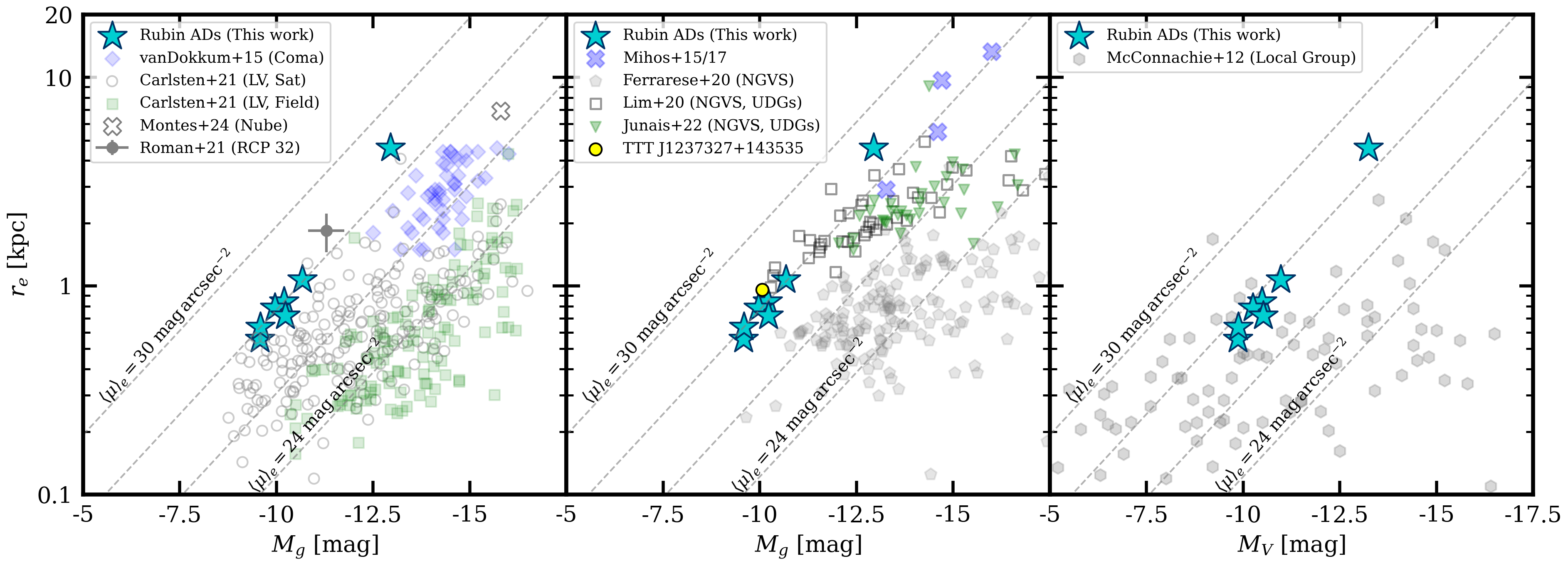}}
      \caption{\textit{Left}: the $g$ band absolute magnitude and effective radius of the seven ADs discovered in Rubin DP2 (marked as stars), compared to UDGs in the Coma Cluster (\citealt{2015ApJ...798L..45V}; diamonds), satellite (open circles) and field (squares) dwarf galaxies within Local Volume (D $\leq 12 \, \Mpc{}$, \citealt{Carlsten2021ApJ...922..267C}), and to the other two ADs Nube (\citealt{Nube_2024A&A...681A..15M}; cross) and RCP~32 (\citealt{2021A&A...656A..44R}; filled circle). \textit{Middle}: as Left panel, compared to other objects in the Virgo Cluster from different surveys: galaxies with \texttt{certain} membership within $0.3 \, \Mpc{}$ from M87 (\citealt{2020ApJ...890..128F}; NGVS, pentagons), UDGs in \citealt{2022A&A...667A..76J} (NGVS, triangles), and in \citealt{2020ApJ...899...69L} (NGVS, squares), VLSB-A/B/C/D galaxies in the Burrell Schmidt Deep Virgo Survey (\citealt{Mihos_2015, Mihos_2017}; blue crosses), and the recently discovered AD TTT~J1237327+143535 (\citealt{TTT_AD}; yellow circle). \textit{Right}: the $V-$band absolute magnitude -- effective radius of these ADs, in comparison to galaxies in the Local Group (D $\leq 3 \, \Mpc{}$)  detected with the star-counting technique (\citealt{McConnachie_2012}, hexagons). The gray dashed lines represent the corresponding effective surface brightness at each magnitude -- radius. The AD~7 candidate is clearly distinguished from the rest of the sample with its effective radius of $4.6\, \kpc{}$. Error bars of the measurements for Rubin ADs are so small as to be seen in this plot.}
    \label{fig:literature_comparison}
\end{figure*}

%%%%%%%%%%%%%%%%%%%%%%%%%%%%%%%%%%%%%%%%%%%%%%%%%%%%%%%%%%%%%%
\section{Physical properties of the seven almost-dark galaxies}\label{section: analysis and results}

    To characterise the physical properties of the seven ADs and mitigate the effect of the Rubin pipeline's small-scale background subtraction on LSB emission, we used \texttt{deep\_coadd.apply\_background(None)}. For each AD, we extract regions of $1.5' \times 1.5'$ centred on the galaxy, with the exception of AD~7 — the most extended object in our sample — for which we use a larger region of $5' \times 5'$. Following the analysis steps in \citet{TTT_AD}, we used the source detection tool \texttt{MTO2} \citep{MTO2} to detect and mask sources around the galaxies, and used the multiband Sérsic fitting tool \texttt{GALFITM} \citep{galfit2010AJ....139.2097P,10.1111/j.1365-2966.2012.20619.x,2013MNRAS.430..330H} to fit Sérsic model on the three bands simultaneously and determine the centre, Sérsic index, position angle, and axis ratio of each galaxy (Fig.~\ref{fig:AD1_galfitm} to Fig.~\ref{fig:AD6_galfitm}). For AD~7, the single Sérsic profile does not fit well, so we used the core-Sérsic model (introduced in \citealt{Graham2003AJ} and \citealt{Trujillo_2004}) with \texttt{IMFIT} \citep{2015ApJ...799..226E} to get its best-fit values (Fig.~\ref{fig:AD7_galfitm}).
    
    We computed the surface brightness profile in the $g$, $r$, and $i$ bands (ADs are not visible in the $u$ band). We applied Galactic extinction correction, following \citet{1998ApJ...500..525S} and \citet{2011ApJ...737..103S}. We then followed \citet{Bakos2008ApJ...683L.103B} to obtain the surface mass density profile, assuming a Kroupa initial mass function \citep{2001MNRAS.322..231K}, and the total stellar mass for each AD. A summary of the physical properties of the ADs is in Tab.~\ref{tab:dwarfs properties} and Tab.~\ref{tab:dwarfs properties2}, and for AD~7 in Tab.~\ref{tab:AD7_properties}.
    
    AD~1 to AD~6 have very flat, core profiles with Sérsic indices $n < 1$. Their effective radii range between $7'' - 15''$, corresponding to $0.6 - 1.1 \, \kpc{}$ at the distance of the Virgo Cluster. They are all extremely faint, with the surface brightness in the $g$ band declining from $26.9 - 27.7 \, \magarcsec{}$ at the centre to $ \sim 31 \, \magarcsec{}$ at distances of two times their effective radii (left panels, Fig. \ref{fig:profiles}, blue squares). The central surface brightness in the $r$ band of these ADs are between $26.4 - 27.1 \, \magarcsec{}$, and declining to $ \sim 30 \, \magarcsec{}$ (orange squares). In the $i$ band, the ADs have $ \mu_{0,i} \sim 26.1 - 26.7 \, \magarcsec{}$, and also declining to $ \sim 29 \, \magarcsec{}$ at their outskirts (gray squares). 

    AD~7 is exceptionally large, with the effective radius $r_\mathrm{e} \sim 58.3 ''$ -- about $4.6\, \kpc{}$. Its profiles follow a core-Sérsic model with a flat core up to $r_\mathrm{b} = 40''$ and joined with a Sérsic profile of $n \sim 0.9$. The integrated half-light radius of AD~7 is at $\sim 4.6\, \kpc{}$. AD~7 is also the faintest one in this sample, with surface brightness at its centre are $28.5, 28.0, 27.8\, \magarcsec{}$, and declining to $ 31.0, 30.0, 29.0 \, \magarcsec{}$ at its outskirts, in $g$-, $r$- and $i$-band, respectively (left panel, last row in Fig. \ref{fig:profiles_continue2}, blue/orange/gray squares represent $g$-, $r$- and $i$-band). 

  The ADs have central surface mass density $ \sim 0.5 - 1 \, \msun \, \pc{}^{-2}$ and declines to $0.01 \, \msun \, \pc{}^{-2}$ at their outskirts (Right panels, Fig. \ref{fig:profiles}). Their total stellar masses range from $9 \times 10^5$ to $ 3 \times 10^6 \, \msun$. An exception is AD~7, which has a central surface mass density of only $\sim 0.25 \, \msun \pc{}^{-2}$ and a total stellar mass of $\sim 10^7 \, \msun$ (last row, right panel, Fig. \ref{fig:profiles}). All the ADs have elliptical morphology ($b/a>= 0.6$), without signatures of being tidally distorted, suggesting those objects are in a relaxed stage. They have mean colours of $(g-r)_0 \sim 0.5$ within their effective radii, likely representative of old, metal-poor stellar populations (see e.g. \citealt{2017MNRAS.468.4039R}).

%%%%%%%%%%%%%%%%%%%%%%%%%%%%%%%%%%%%%%%%%%%%%%%%%%%%%%%%%%%%%%
\section{Discussion}

All seven ADs are most likely associated with the Virgo Cluster for two reasons. First, they have projected distances between $0.2-1.1 \, \Mpc{}$ from M49 -- the centre of the Virgo B subcluster with $r_\mathrm{vir}\sim 1 \, \Mpc{}$ \citep{2012ApJS..200....4F} -- placing them within the virial radius of the Virgo B subcluster (except for AD~4 at 1.1 Mpc, which is near the edge of that virial radius). Second, their extended effective radii ($r_\mathrm{e} \sim 7''-15''$ in the cases of AD~1 to AD~6; and even $58''$ for AD~7) imply that they are likely at the distance of the Virgo Cluster (where their physical sizes are $\sim 0.6-4.6\, \kpc{}$), which is consistent with known low-mass dwarfs in Virgo. If these ADs were at twice the Virgo distance ($\sim 30 \, \Mpc{}$), their physical sizes would be twice as large, which makes them extremely large and outside the regime that simulations of galaxy formation predict \citep{2022NatAs...6..897S}. Especially AD~7, if at $30 \, \Mpc{}$, would have $r_\mathrm{e} \sim 8 \, \kpc{}$ and $M_g \sim -14 \, \mathrm{mag} $, making it a giant object in an unrealistic region of the effective radius -- magnitude plane. Their smooth, unresolved surface brightnesses also place a lower bound on the distance: at $\sim 5 \, \Mpc{}$ the TRGB ($M_i \sim -4 \, \mathrm{mag}$, \citealt{1993ApJ...417..553L}) would reach $m_i \sim 24.5 \, \mathrm{mag}$, detectable above the $5\sigma$ limit of $m_i \sim 25.4 \, \mathrm{mag}$, yet no individual stars are seen. This rules out the possibility that these ADs are foreground galaxies.

Six ADs (AD~1 to AD~6) are similar in effective radius to AD TTT~J1237327+143535 \citep{TTT_AD}. Compared to other dwarf galaxies in the Local Volume (Fig.~\ref{fig:literature_comparison}, Left panel), they are at the very faint end and larger than most of the galaxies in the same magnitude range. Compared to galaxies and UDGs in the Virgo Cluster (Middle panel), they are among the faintest ones. Compared to satellites in the Local Group (Right panel), these ADs are analogues of those which are at the faintest end of that sample, such as And~XXI, And~XXIII, and And~XXV.

The candidate AD~7 is the flattest, most extended, extremely faint object in this sample. On the magnitude -- effective radii plane, AD~7 also stands out as the largest AD in this sample, separate from the rest with its $r_\mathrm{e} \sim 4.6 \, \kpc{}$ at the distance of the Virgo Cluster (Fig.~\ref{fig:literature_comparison}). It is the faintest galaxy in total magnitude compared to galaxies at the same effective radii, and the largest in effective radii compared to galaxies of similar magnitude. Herschel 250 micron and Planck 857 GHz at the position of AD~7 both show no detection of Galactic cirrus. Also, Galactic extinction is $E(B-V) = 0.02$ at that position. Hence, it is less likely that AD~7 is part of Galactic cirrus. Besides, AD~7 is located near many bright objects, appearing to be in an oversubtracted area, in which its flux can be underestimated. The significant uncertainties in its surface brightness profiles may reflect high sky background brightnesses. Follow-up observations with an LSB-friendly observing strategy are needed to verify if this object is real. If confirmed, AD~7 would be one of the most extended and the faintest LSBs discovered in the Virgo Cluster (compared to \citealt{Mihos_2015, Mihos_2017} and to \citealt{2020ApJ...899...69L}), as well as compared to those in the Local Group and the Local Volume. 

None of the seven ADs have an HI counterpart in the ALFALFA survey \citep{2018ApJ...861...49H}, yielding an upper limit on their HI mass of $10^7 \, \msun$ at the Virgo distance. Except for AD~4, which lies outside the Virgo Environmental Survey Tracing Ionised Gas Emission survey (VESTIGE) footprint, the other ADs have no detection in H$\alpha$ from VESTIGE \citep{2018A&A...614A..56B}, indicating an absence of ongoing star formation. Such red, low-surface-brightness, low-mass galaxies are expected to be HI-poor and have no star-forming activity. This picture is consistent with TTT~J1237327+143535, which shows the same HI non-detection and red, quiescent properties.

The discovery of seven AD candidates in Rubin DP2 data demonstrates that this extreme population can be systematically identified in wide-field survey imaging. It is worth emphasizing that these candidates have been missed in other deep surveys of the Virgo Cluster. 
While this work does not present a complete census of ADs in the surveyed field, the objects reported here were identified through visual inspection and are intended as a proof-of-concept demonstration — establishing the existence of this class of object within the Rubin DP2 data and illustrating the survey's capacity to detect such extremely low-surface-brightness systems. Additional candidates almost certainly reside within the same field, and a dedicated systematic, automated search across the full DP2 footprint — and ultimately the 10-year Rubin LSST of $ 18,000 \deg^2$ coverage — will be required to build a statistically complete sample in Virgo and beyond. 
These ADs populate the very faint end of the galaxy luminosity function, where the predicted abundance of dark matter haloes from different dark matter models shows discrepancies. By accurately determining their abundance, we can rule out different dark matter models and improve our understanding of the nature of dark matter \citep{2021JCAP...08..062N, 2024ApJ...967...61N}.

Two sources of systematic uncertainty affect our measurements. First, the fluxes at the outskirts of these ADs could be underestimated because of the visit-level background subtraction ($128$ pixel spline) in Rubin DP2 (as discussed in \citealt{10.1093/mnras/stad180, 2024MNRAS.528.4289W}). Therefore, the shape of their surface brightness profiles, as well as their total magnitudes, are likely underestimated. 
Second, the formal uncertainties returned by \texttt{GALFITM} (or \texttt{IMFIT}) are likely to underestimate the true uncertainties in the fitted structural parameters (\citealt{2007ApJS..172..615H}, \citealt{10.1111/j.1365-2966.2010.17855.x}), as they do not fully account for systematic effects such as uncertainties in the sky background, PSF modelling, and masking.

%%%%%%%%%%%%%%%%%%%%%%%%%%%%%%%%%%%%%%%%%%%%%%%%%%%%%%%%%%%%%%
\section{Conclusions}\label{section:conclusions}

    We report the finding of a population of seven almost-dark galaxies around M49 in the Virgo Cluster with Rubin LSST~DP2. These ADs have a central surface brightness of $ 26.9 - 28.5 \, \magarcsec{}$ in the $g$-band, flat profiles with Sérsic index $n < 1$, effective radii of $0.6 - 4.6\, \kpc{}$, and stellar masses from $10^6 \, \msun{}$ to $10^7 \, \msun{}$. Finding these new objects in Rubin~LSST~DP2 data shows the usefulness of the Rubin~LSST survey in studying low-surface brightness objects. 
    
    The full 10-year Vera C. Rubin Observatory LSST survey, with its combination of depth and sky coverage, will deliver a more complete sample of this extreme population surrounding the Virgo Cluster - complementing other deep surveys of this cluster - and across vast areas of the sky. The resulting statistical sample of ADs will be transformative for constraining the faint-end of the galaxy luminosity function and discriminating between dark matter models.

%%%%%%%%%%%%%%%%%%%%%%%%%%%%%%%%%%%%%%%%%%%%%%%%%%%%%%%%%%%%%%
\begin{acknowledgements}
      MNL thanks her supervisor, Ignacio Trujillo, for the discovery of the objects presented in this work and for invaluable scientific guidance throughout this project. This work is co-funded by the European Union (MSCA Doctoral Network EDUCADO, GA 101119830, Widening Participation, ExGal-Twin, GA 101158446, and UNDARK, GA 101159929). This work used the resources of the Rubin US Data Facility hosted by the SLAC Shared Science Data Facility (S3DF) at SLAC National Accelerator Laboratory, funded by Department of Energy under Contract No. DE-AC02-76SF00515. This research uses services or data provided by the Rubin Science Platform at NSF-DOE Vera C. Rubin Observatory, which is jointly funded by the U.S. National Science Foundation and the U.S. Department of Energy, Office of Science. This work is part of grant CEX2025-001609-S, awarded to the IAC under the Severo Ochoa Centre of Excellence program and funded by 250 MICIU/AEI/10.13039/501100011033. IT acknowledges grant PID2022-140869NB-I00 and IAC project P/302302, financed by MCIN. JHK acknowledges grants PID2022-136505NB-I00, funded by MCIN/AEI/10.13039/501100011033 and EU, ERDF. J acknowledges the Alexander von Humboldt Foundation through a Humboldt Research Fellowship. AEW acknowledges support from the Science and Technology Facilities Council (STFC; grant number ST/X001318/1). R.D. acknowledges support by the ANID BASAL project FB210003.
\end{acknowledgements}

\bibliographystyle{bibtex/aa}
\bibliography{references.bib}

\begin{appendix}
%%%%%%%%%%%%%%%%%%%%%%%%%%%%%%%%%%%%%%%%%%%%%%%%%%%%%%%%%%%%%%%
\onecolumn
\section{Properties of $7$ almost-dark galaxies}

\begin{figure*}[ht!]
    \centering

    \begin{subfigure}{\textwidth}
        \includegraphics[width=0.96\linewidth]{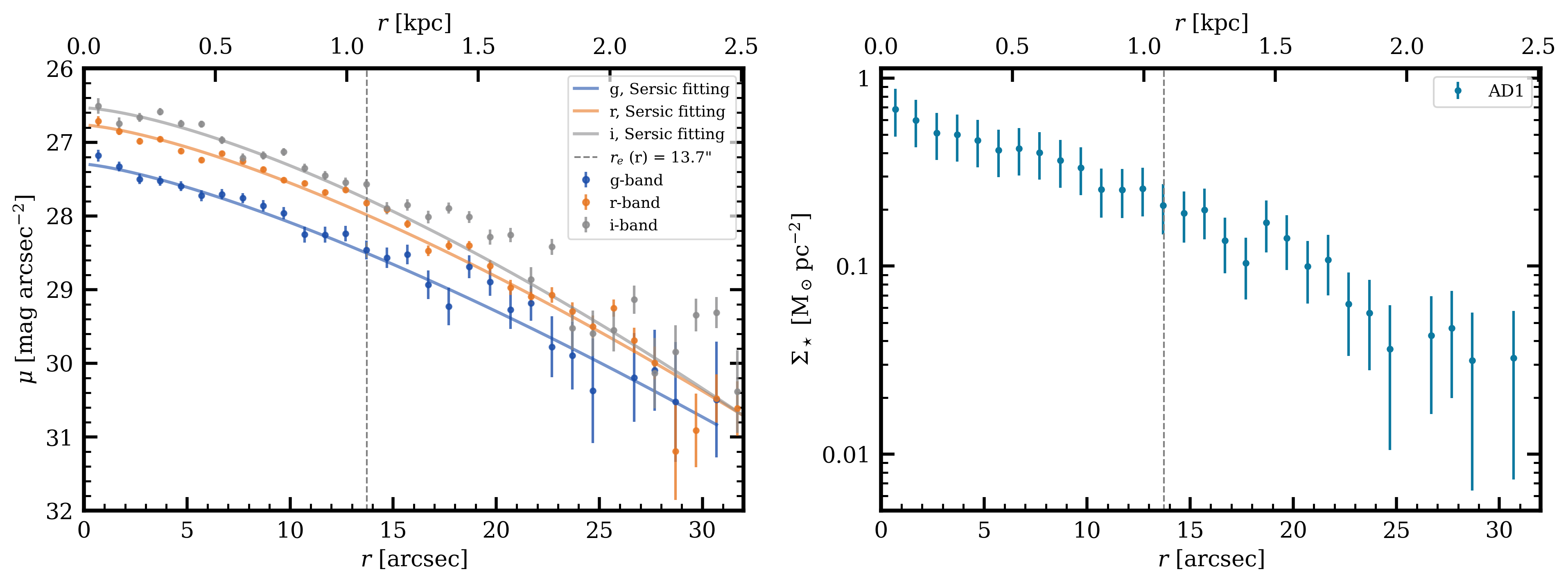}
    \end{subfigure}

    \begin{subfigure}{\textwidth}
        \includegraphics[width=0.96\linewidth]{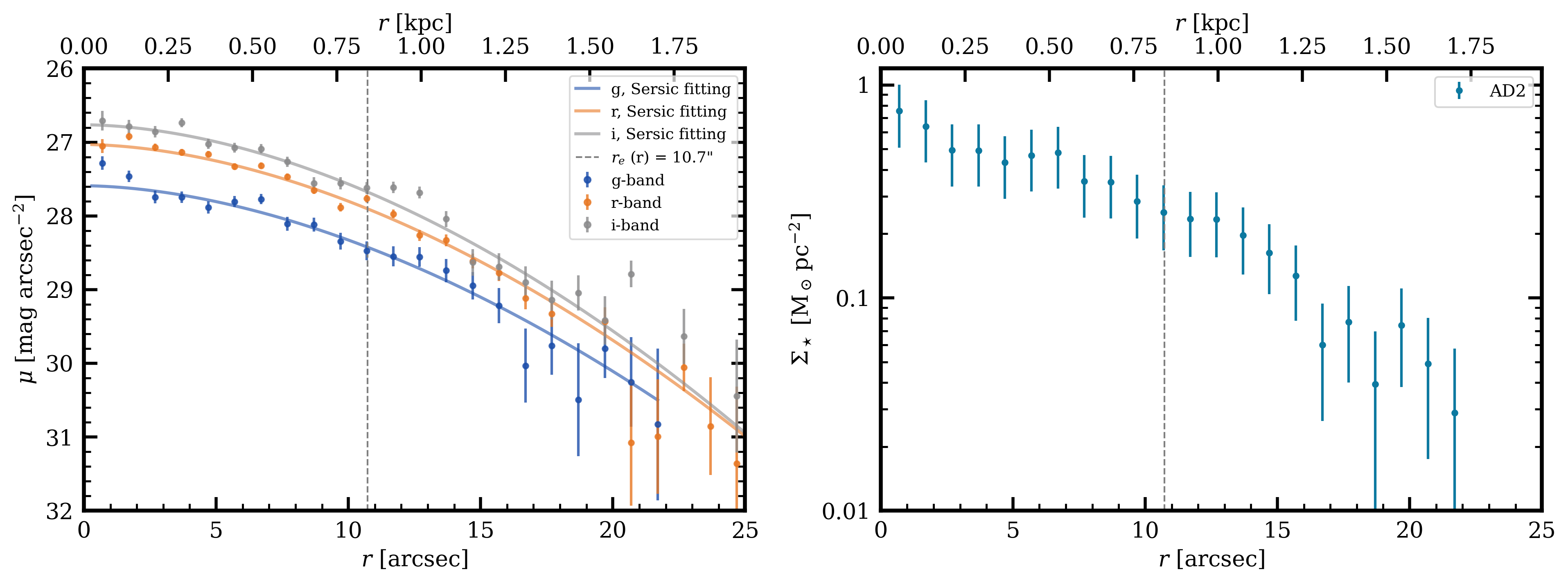}
    \end{subfigure}

    \begin{subfigure}{\textwidth}
        \includegraphics[width=0.96\linewidth]{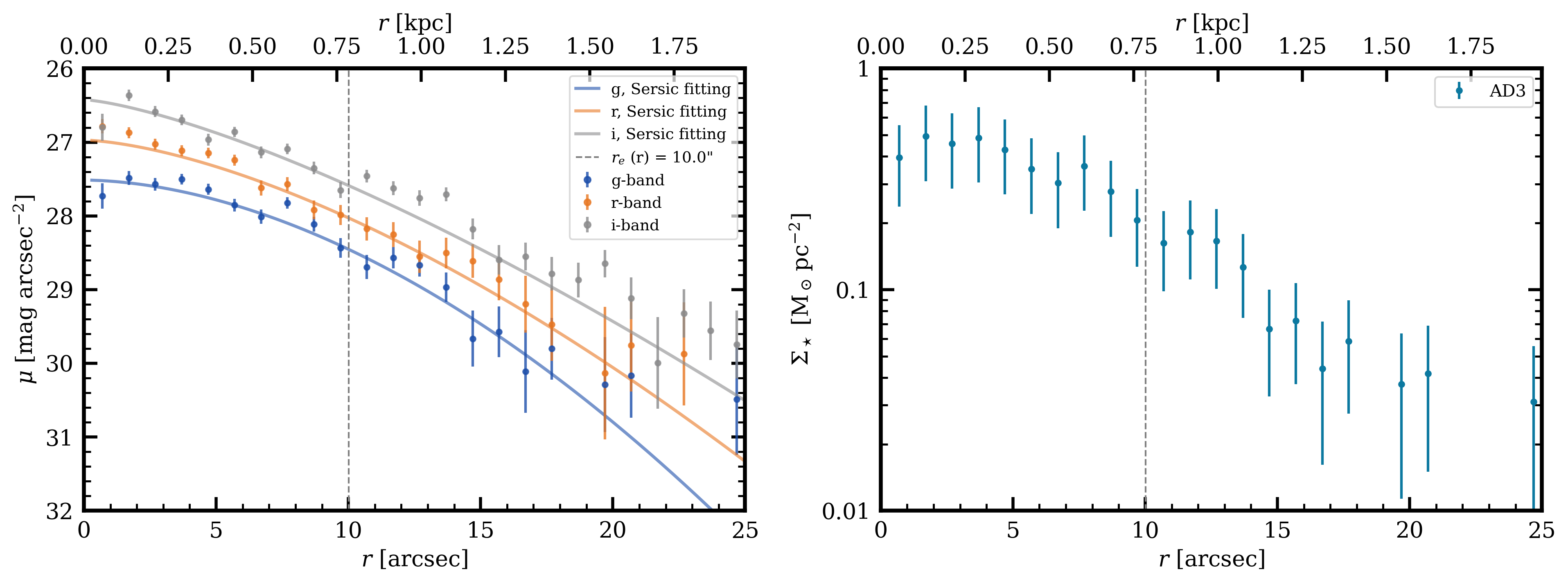}
    \end{subfigure}
    \caption{\textit{Left panel}: Surface brightness profiles in $g-$ (blue points), $r-$ (orange points) and $i-$ (gray points) bands and their best-fit 2D Sérsic (lines) of a AD. \textit{Right panel}: surface mass density profiles of that AD. Here, from top to bottom, we show Rubin~J122848.6+085727 (AD~1), Rubin~J123148.4+090928 (AD~2), Rubin~J123558.9+084825 (AD~3), with vertical lines showing the effective radii of each AD (continued on next pages).}
    \label{fig:profiles}
    
\end{figure*}

\begin{figure*}\ContinuedFloat
    \centering
    \begin{subfigure}{\textwidth}
        \includegraphics[width=0.95\linewidth]{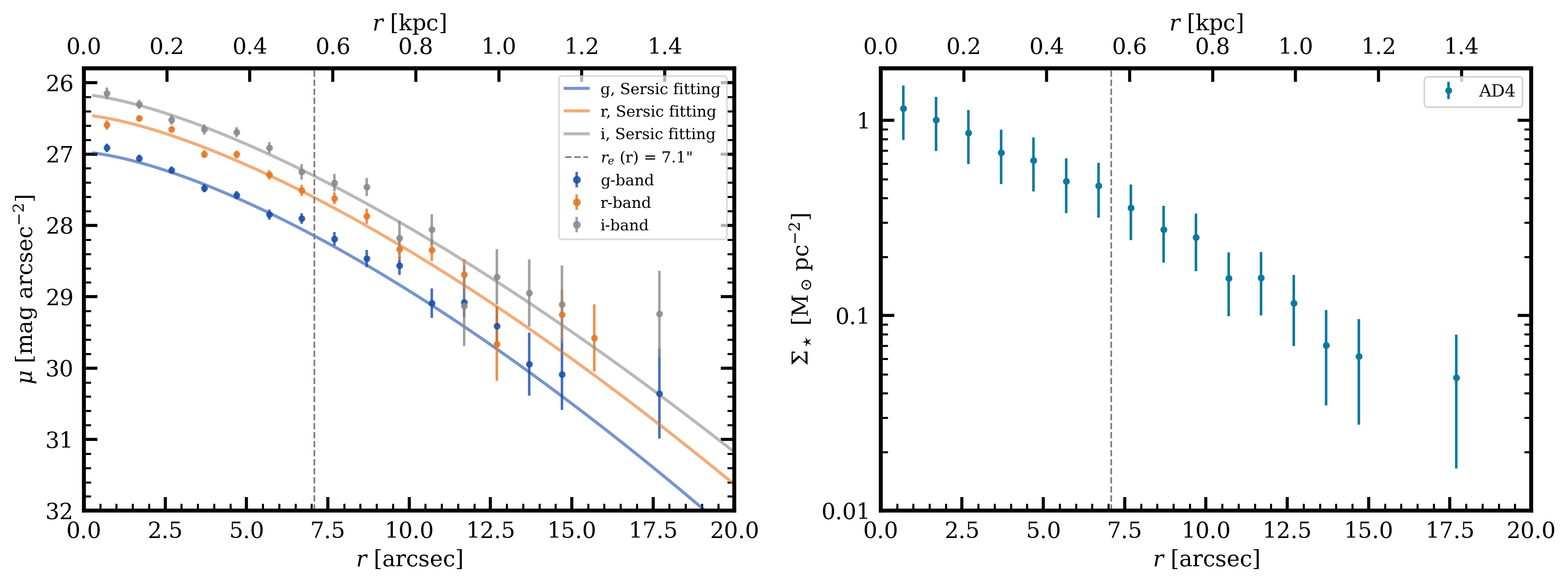}
    \end{subfigure}

    \begin{subfigure}{\textwidth}
        \includegraphics[width=0.95\linewidth]{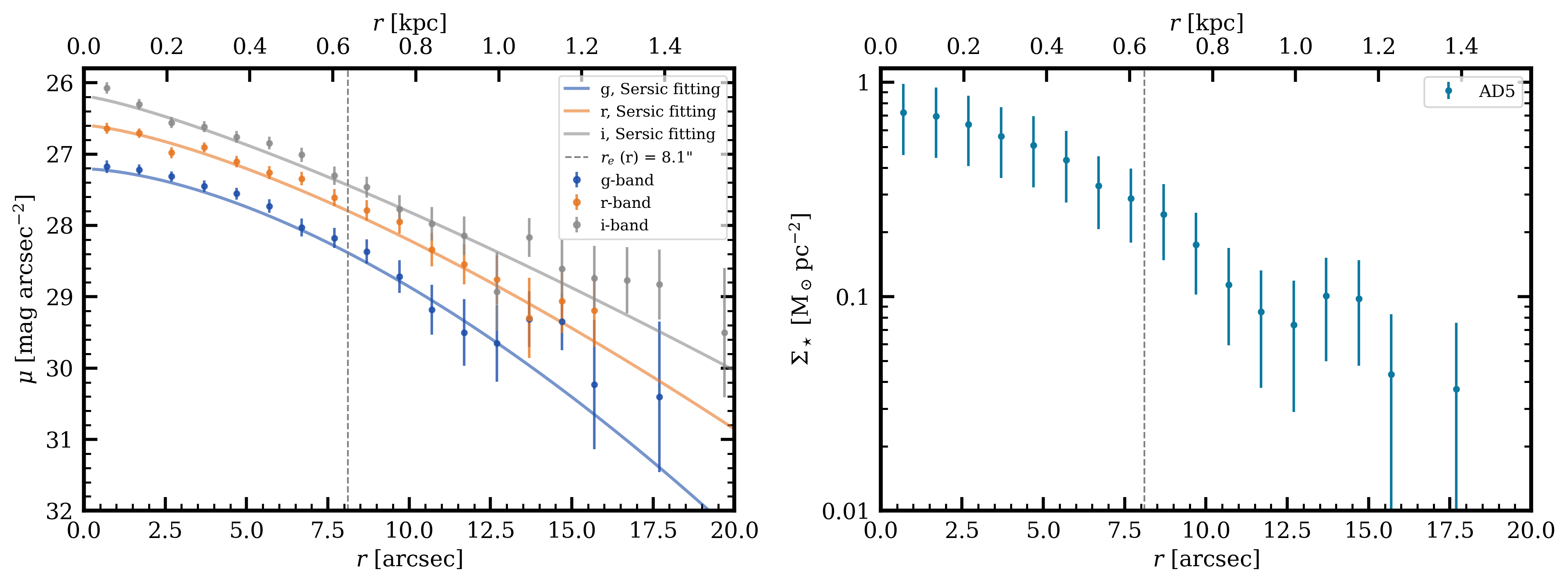}
    \end{subfigure}

    \begin{subfigure}{\textwidth}
        \includegraphics[width=0.95\linewidth]{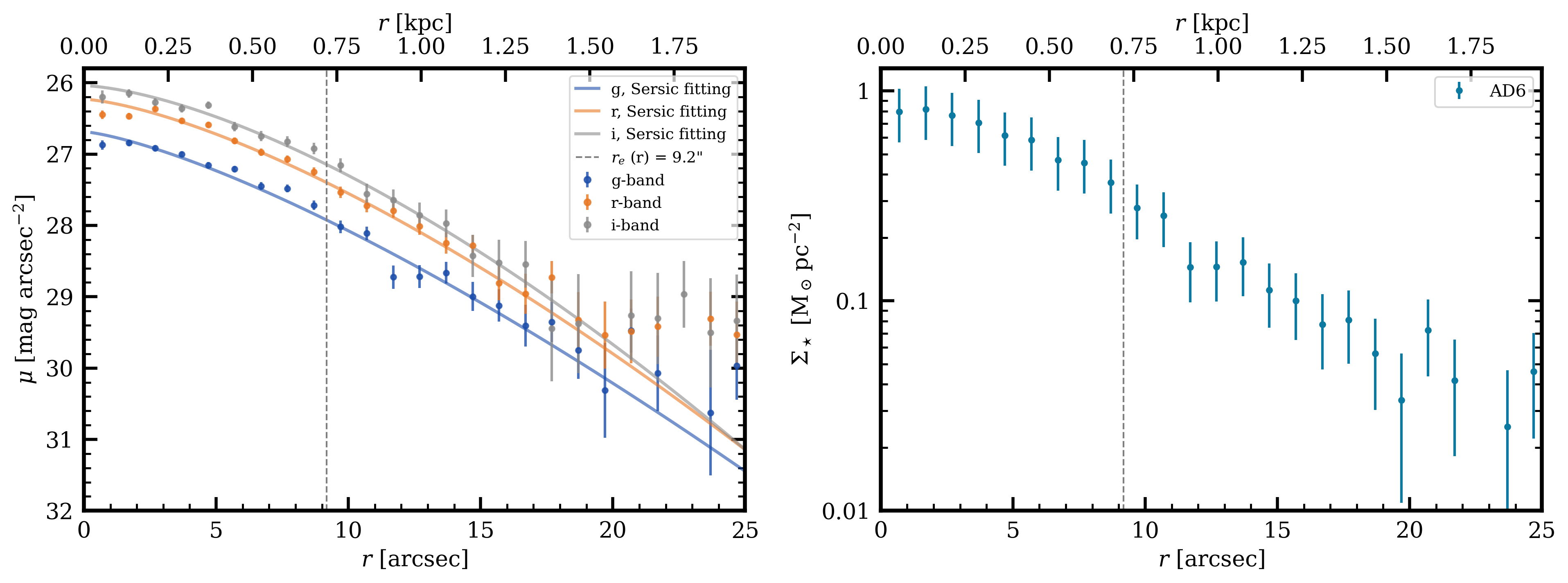}
    \end{subfigure}
    \caption{Continued, for Rubin~J121723.8+053701 (AD~4), Rubin~J122659.3+090234 (AD~5) and Rubin~J123233.1+082852 (AD~6).}
    \label{fig:profiles_continue}
\end{figure*}

\begin{figure*}\ContinuedFloat
    \centering
    \begin{subfigure}{\textwidth}
        \includegraphics[width=\linewidth]{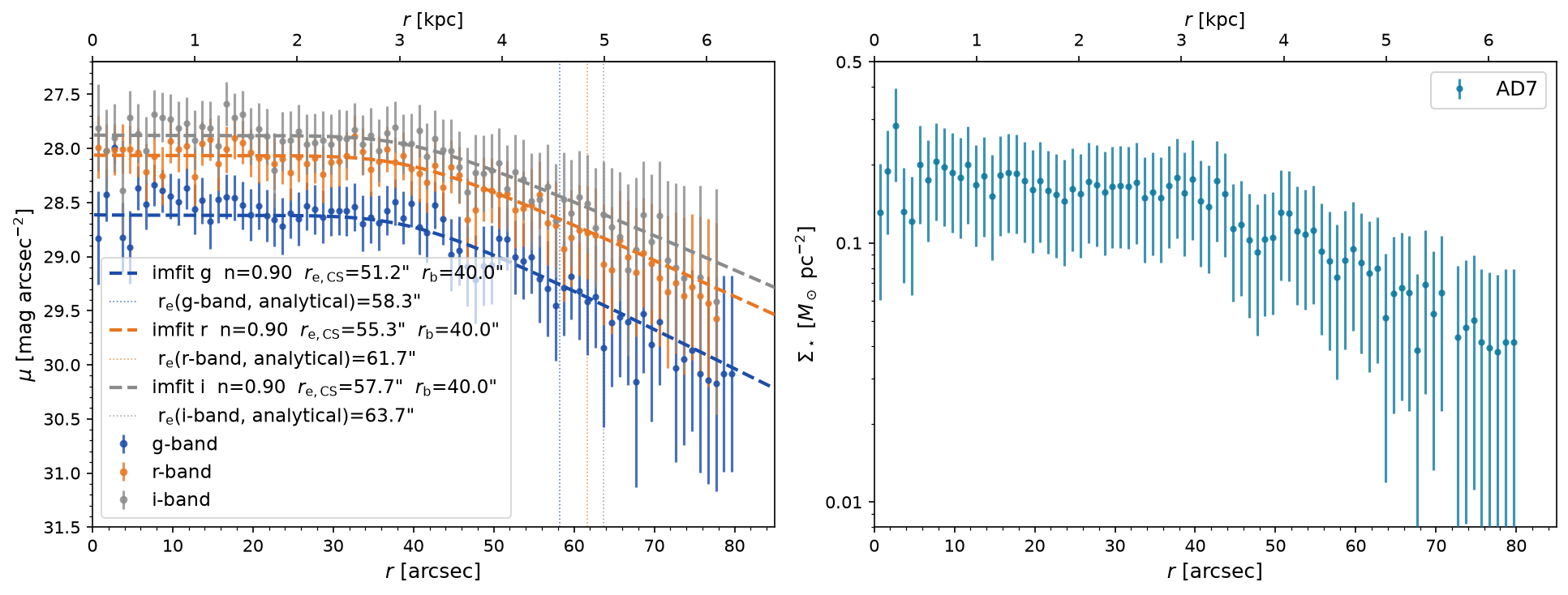}
    \end{subfigure}

    \caption{Continued, for Rubin~J122703.1+080018 (AD~7). Vertical lines show the analytical half-light radius from the core-Sérsic model in each band.}
    \label{fig:profiles_continue2}
\end{figure*}

\FloatBarrier %\usepackage{placeins}
\clearpage

% Ngoc: need to input new value for AD2-AD7 and <mu_>e for AD1, AD2
\begin{sidewaystable*}[ht!]
\centering
\caption{ Properties\tablefoottext{a}{} of $7$ almost-dark galaxies around M49 in Rubin LSST DP2. Shows AD~1, AD~2, and AD~3. Continued on the next page (Tab.\ref{tab:dwarfs properties2}).  }
\begin{tabular}{lccc}
\hline\hline \\[0.01cm]
   & Rubin~J122848.6+085727 (AD~1) & Rubin~J123148.4+090928 (AD~2) & Rubin~J123558.9+084825 (AD~3) \\[0.01cm] 
   \hline 
 RA [$^\circ$] & $187.202634$ & $187.951943$ & $188.995572$\\ 
 DEC [$^\circ$] & $8.957567$ &  $9.157953$ & $8.807054$\\
 $r_\mathrm{e}$ [$''$] ($g$ band) & $14.33\pm0.39$ & $11.07\pm0.25$  & $9.52\pm0.25$\\
 $r_\mathrm{e}$ [$''$]  ($r$ band) & $13.72 \pm 0.24$ & $10.73\pm0.17$  &$10.01\pm0.18$\\
 $r_\mathrm{e}$ [$''$]  ($i$ band) & $13.16\pm0.38$ & $10.41\pm0.27$  & $10.46\pm0.34$ \\
 $r_\mathrm{e}$ [$\kpc{}$] ($g$ band) & $1.12\pm0.03$ & $0.87\pm0.02$  & $0.75\pm0.02$\\
 $r_\mathrm{e}$ [$\kpc{}$]  ($r$ band) & $1.08\pm0.02$ & $0.84\pm0.01$  &$0.79\pm0.01$\\
 $r_\mathrm{e}$ [$\kpc{}$]  ($i$ band) & $1.03\pm0.03$ & $0.82\pm0.02$  & $0.82\pm0.03$ \\
 $n$ ($g$ band) &  $0.75\pm 0.03$ & $0.56\pm0.02$ & $0.55\pm0.03$\\
 $n$  ($r$ band) &  $0.70\pm 0.03$ & $0.56\pm0.01$ & $0.65\pm0.02$\\
 $n$ ($i$ band) &  $0.75\pm 0.03$ & $0.56\pm0.02$ & $0.73\pm0.03$\\
 $b/a$ & $0.83\pm0.01$ &  $0.91\pm0.01$ & $0.91\pm0.01$ \\ 
 PA [$^\circ$] &  $63.7\pm2.8$ &  $-21.74\pm6.15$ & $51.01\pm7.89$\\
 $\mu_{0,g}$ [$\magarcsec$]  & $27.18\pm0.08$& $27.28\pm0.09$ & $27.73\pm0.17$\\
 $\mu_{0,r}$ [$\magarcsec$]  & $26.71\pm0.07$ & $27.05\pm0.09$ & $26.78\pm0.10$\\
 $\mu_{0,i}$ [$\magarcsec$]  & $26.51\pm0.11$ & $26.71\pm0.13$ & $26.60\pm 0.18$\\
 $\left\langle\mu_{g}\right\rangle_\mathrm{e}$ [$\magarcsec$] & $28.01 \pm 0.07$ & $28.02\pm0.06$ & $27.93\pm0.07$\\
 $\left\langle\mu_{r}\right\rangle_\mathrm{e}$ [$\magarcsec$] & $27.43\pm0.05$ & $27.46\pm0.04$ & $27.52\pm0.05$\\
 $\left\langle\mu_{i}\right\rangle_\mathrm{e}$ [$\magarcsec$] & $27.16\pm0.07$ & $27.19\pm0.06$ & $27.11\pm0.08$\\
 $(g-r)_0$ \tablefoottext{b}{} &  $0.50\pm0.03$ & $0.54\pm 0.05$ & $0.49 \pm 0.06$\\
 $g$ [mag]  & $20.43\pm0.03$ & $20.91\pm0.03$ & $21.15\pm0.03$\\
 $r$ [mag] & $19.95\pm 0.02$ & $20.41\pm0.02$ & $20.63\pm0.02$\\
 $i$ [mag] & $19.77\pm 0.04$ & $20.20\pm0.03$ & $20.12\pm0.04$\\
 $M_\star$ $[10^6 \, \mathrm{M_\odot}]$ &  $2.1\pm0.2$ & $1.6\pm0.2$ & $1.2\pm0.2$ \\
\hline
\end{tabular}
\label{tab:dwarfs properties}
\tablefoot{\tablefoottext{a}{The effective radius $r_\mathrm{e}$, Sérsic index $n$, axis ratio $b/a$, position angle PA, and total magnitude in all three bands are derived from the fitting using \texttt{GALFITM}.} The central and mean effective surface brightness values are corrected for Galactic extinction.
\tablefoottext{b}{Median Galactic extinction corrected $(g-r)_0$ colour within $r_\mathrm{e}$ of the galaxies and bootstrap errors.}}
   
\end{sidewaystable*}

\begin{sidewaystable*}
\centering
\caption{ Same as Tab.\ref{tab:dwarfs properties}, for AD~4, AD~5 and AD~6.}
\begin{tabular}{lccc}
\hline\hline \\[0.01cm]
   & Rubin~J121723.8+053701 (AD~4) & Rubin~J122659.3+090234 (AD~5) & Rubin~J123233.1+082852 (AD~6)  \\[0.01cm] 
   \hline 
 RA [$^\circ$] & $184.349420$& $186.747124$ & $188.138187$\\ 
 DEC [$^\circ$] & $5.617167$ &  $9.042990$ & $8.481132$ \\
 $r_\mathrm{e}$ [$''$] ($g$ band) & $6.92\pm0.14$ & $7.29\pm0.17$  & $9.35\pm0.19$ \\
 $r_\mathrm{e}$ [$''$] ($r$ band) & $7.09\pm0.10$ & $8.11\pm0.14$  & $9.18\pm0.13$\\
 $r_\mathrm{e}$ [$''$] ($i$ band) & $7.24\pm0.16$ & $8.86\pm0.25$  & $9.03\pm0.19$ \\
 $r_\mathrm{e}$ [$\kpc{}$] ($g$ band) & $0.54\pm0.01$ & $0.57\pm0.01$  & $0.73\pm0.02$\\
 $r_\mathrm{e}$ [$\kpc{}$] ($r$ band) & $0.56\pm0.01$ & $0.64\pm0.01$  & $0.72\pm0.01$\\
 $r_\mathrm{e}$ [$\kpc{}$] ($i$ band) & $0.57\pm0.01$ & $0.69\pm0.02$  & $0.71\pm0.02$ \\
 $n$ ($g$ band) &  $0.68\pm0.03$ & $0.61\pm0.03$ & $0.74\pm0.02$ \\
 $n$ ($r$ band) &  $0.69\pm0.02$ & $0.71\pm0.02$ & $0.70\pm0.01$ \\
$n$ ($i$ band) &  $0.70\pm0.04$ & $0.80\pm0.04$ & $0.66\pm0.02$ \\

 $b/a$ & $0.87\pm0.01$ &  $0.88\pm0.01$ & $0.74\pm0.01$ \\ 
 PA [$^\circ$] & $79.93\pm4.50$ &$-55.71\pm 4.84$ &$45.27\pm1.35$\\
 $\mu_{0,g}$ [$\magarcsec$]  & $26.91\pm0.06$& $27.18\pm0.09$ & $26.87\pm0.06$ \\
 $\mu_{0,r}$ [$\magarcsec$]  & $26.59\pm0.07$ & $26.64\pm0.08$ & $26.45\pm0.06$ \\
 $\mu_{0,i}$ [$\magarcsec$]  & $26.15\pm0.08$ & $26.07\pm 0.08$ & $26.20\pm0.10$\\
 $\left\langle\mu_{g}\right\rangle_\mathrm{e}$ [$\magarcsec$] & $27.58\pm0.05$ & $27.71\pm0.06$ & $27.39\pm0.05$ \\
 $\left\langle\mu_{r}\right\rangle_\mathrm{e}$ [$\magarcsec$] & $27.08\pm0.04$ & $27.25\pm0.04$ & $26.86\pm 0.03$ \\
 $\left\langle\mu_{i}\right\rangle_\mathrm{e}$ [$\magarcsec$] & $26.80\pm0.05$ & $26.98\pm0.07$ & $26.61\pm0.05$ \\
 $(g-r)_0$  &  $0.56\pm 0.04$ & $0.51\pm0.06$ & $0.47\pm0.03$\\
 $g$ [mag]  & $21.54\pm0.02$ & $21.54\pm0.03$ & $20.87\pm0.02$\\
 $r$ [mag] & $20.98\pm0.02$ & $20.85\pm0.02$ & $20.38\pm0.01$\\
 $i$ [mag] & $20.66\pm0.02$ & $20.39\pm0.03$ & $20.17\pm0.02$\\
 $M_\star$ $[10^6 \, \mathrm{M_\odot}]$ &  $1.0\pm0.1$ & $0.9\pm0.1$ & $1.2\pm0.1$ \\
\hline
\end{tabular}
\label{tab:dwarfs properties2}
\end{sidewaystable*} 

\begin{table}
\centering
\caption{Physical properties of AD~7.}
\begin{tabular}{lc}
\hline\hline \\[0.01cm]
   & Rubin~J122703.1+080018  (AD~7)  \\[0.01cm] 
   \hline 
 RA [$^\circ$] & $186.763211$\\ 
 DEC [$^\circ$] & $8.005028$\\
 $r_\mathrm{e,CS}$ [$''$] ($g$ band) &$51.3\pm0.8$\\
 $r_\mathrm{e,CS}$ [$''$] ($r$ band)  &$55.4\pm0.7$\\
 $r_\mathrm{e,CS}$ [$''$] ($i$ band) & $57.7\pm1.1$\\
 %$r_\mathrm{e,CS}$ [$\kpc{}$] ($g$ band)  &$3.94 \pm 0.03$\\
 %$r_\mathrm{e,CS}$ [$\kpc{}$] ($r$ band) &$3.94 \pm 0.03$\\
 %$r_\mathrm{e,CS}$ [$\kpc{}$] ($i$ band) &$3.94 \pm 0.03$\\
 $r_\mathrm{b}$ [$''$] ($g$ band) &$40.0\pm1.0$\\
 $r_\mathrm{b}$ [$''$] ($r$ band)  &$40.0\pm1.0$\\
 $r_\mathrm{b}$ [$''$] ($i$ band) & $40.0\pm1.0$\\
 %$r_\mathrm{b}$ [$\kpc{}$] ($g$ band)  &$3.94 \pm 0.03$\\
 %$r_\mathrm{b}$ [$\kpc{}$] ($r$ band) &$3.94 \pm 0.03$\\
 %$r_\mathrm{b}$ [$\kpc{}$] ($i$ band) &$3.94 \pm 0.03$\\
 $r_\mathrm{e}$  [$''$] ($g$ band) \tablefoottext{a}{} & $58.3 \pm 0.7$\\
 $r_\mathrm{e}$ [$''$] ($r$ band)  &$61.7\pm 0.5$\\
 $r_\mathrm{e}$ [$''$] ($i$ band) & $63.7\pm 0.9$\\
 $r_\mathrm{e}$ [$\kpc{}$] ($g$ band)  &$4.6 \pm 0.3$\\
 $r_\mathrm{e}$ [$\kpc{}$] ($r$ band) &$4.8 \pm 0.3$\\
 $r_\mathrm{e}$ [$\kpc{}$] ($i$ band) &$5.0 \pm 0.3$\\
 $n$ ($g$ band) &$0.9\pm0.1$\\
 $n$ ($r$ band) &$0.9\pm0.1$\\
 $n$ ($i$ band) &$0.9\pm0.1$\\
 $\gamma$ & $0.0$\\
 $\alpha$ & $9.0$ \\
 $b/a$ & $0.8\pm0.1$ \\ 
 PA [$^\circ$] &$-70.5\pm1.5$ \\
 $\mu_{0,g}$ [$\magarcsec$]  & $28.50\pm0.46$\\
 $\mu_{0,r}$ [$\magarcsec$] & $27.99\pm0.34$\\
 $\mu_{0,i}$ [$\magarcsec$]  & $27.81\pm0.30$\\
 $\left\langle\mu_{g}\right\rangle_\mathrm{e}$ [$\magarcsec$] & $28.80\pm0.03$\\
 $\left\langle\mu_{r}\right\rangle_\mathrm{e}$ [$\magarcsec$]  & $28.22\pm0.02$\\
 $\left\langle\mu_{i}\right\rangle_\mathrm{e}$ [$\magarcsec$]  & $28.06\pm 0.04$\\
 $(g-r)_0$ & $0.5\pm0.1$\\
 $g$ [mag]  & $18.1\pm0.1$\\
 $r$ [mag]  & $17.5 \pm 0.1$\\
 $i$ [mag]  & $17.3\pm0.1$\\
 $M_\star$ $[10^6 \, \mathrm{M_\odot}]$ & $11.8\pm0.9$ \\
\hline
\end{tabular}
\label{tab:AD7_properties}
\tablefoot{We fixed $\gamma$ and $\alpha$ to obtain the core-Sérsic profile in \texttt{IMFIT}; the other parameters of a core-Sérsic profile (effective radii of the core-Sérsic model ($r_\mathrm{e, CS}$), break radii $r_\mathrm{b}$, Sérsic index $n$, position angle PA, and axis ratio b/a) are obtained from the fit. The central and mean effective surface brightness values are corrected for Galactic extinction. We integrated the core-Sérsic model to get the total magnitude in each band.\tablefoottext{a}{The effective radius $r_\mathrm{e}$ is derived from integration of the core-Sérsic model, taking the radii at half-light.}}
\end{table}

\begin{figure*}[ht!]
     \centering
        \includegraphics[width=1\linewidth]{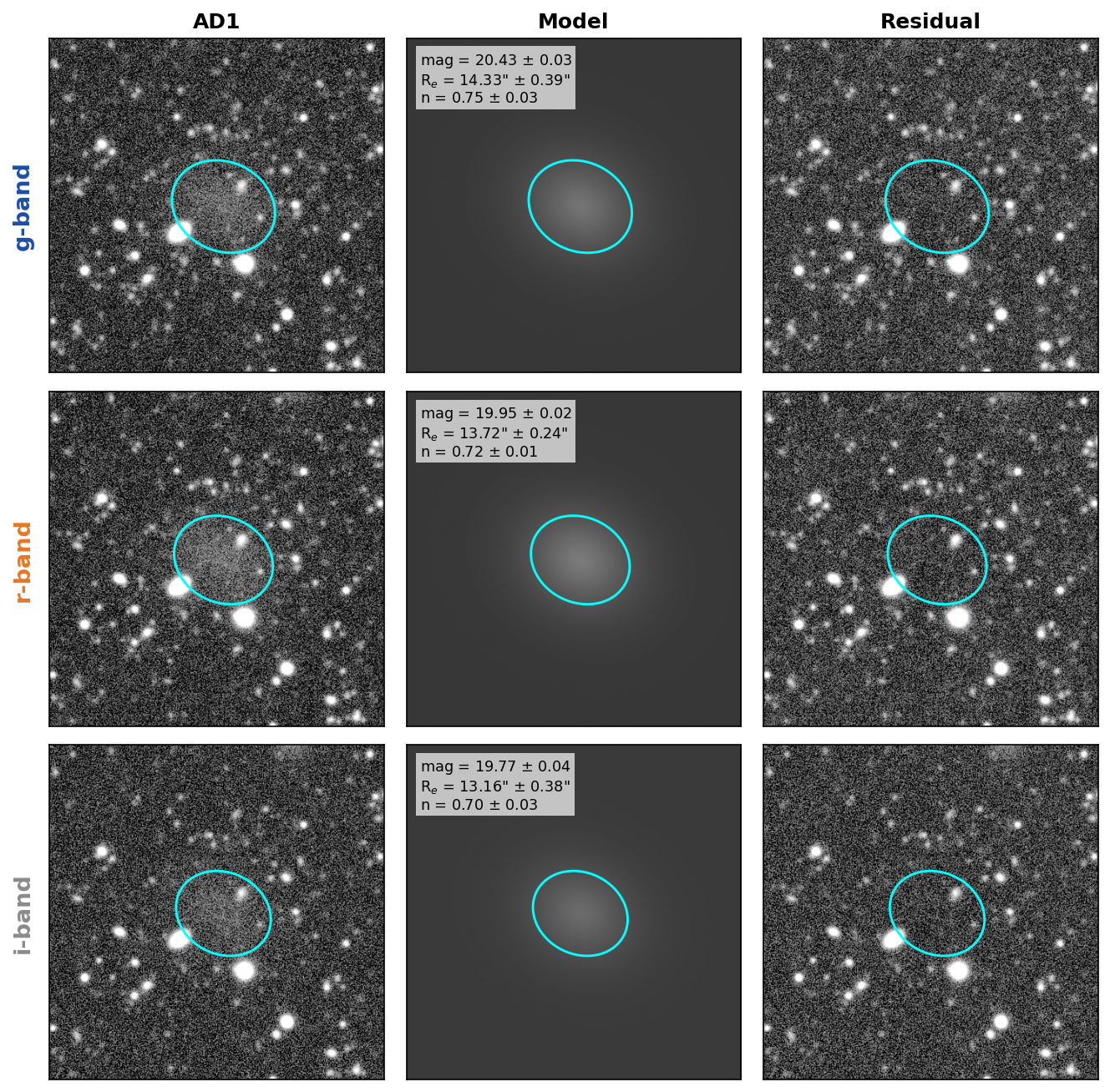}
        \caption{Image of AD~1 in $g$, $r$, and $i$ bands (first column, from top to bottom), compared to their models from \texttt{GALFTIM} (middle column) and the residuals (last column). The ellipses show the effective radii in each band. Continued on the next pages for other ADs.}
        \label{fig:AD1_galfitm}
\end{figure*}
\begin{figure*}[ht!]
     \centering
        \includegraphics[width=1\linewidth]{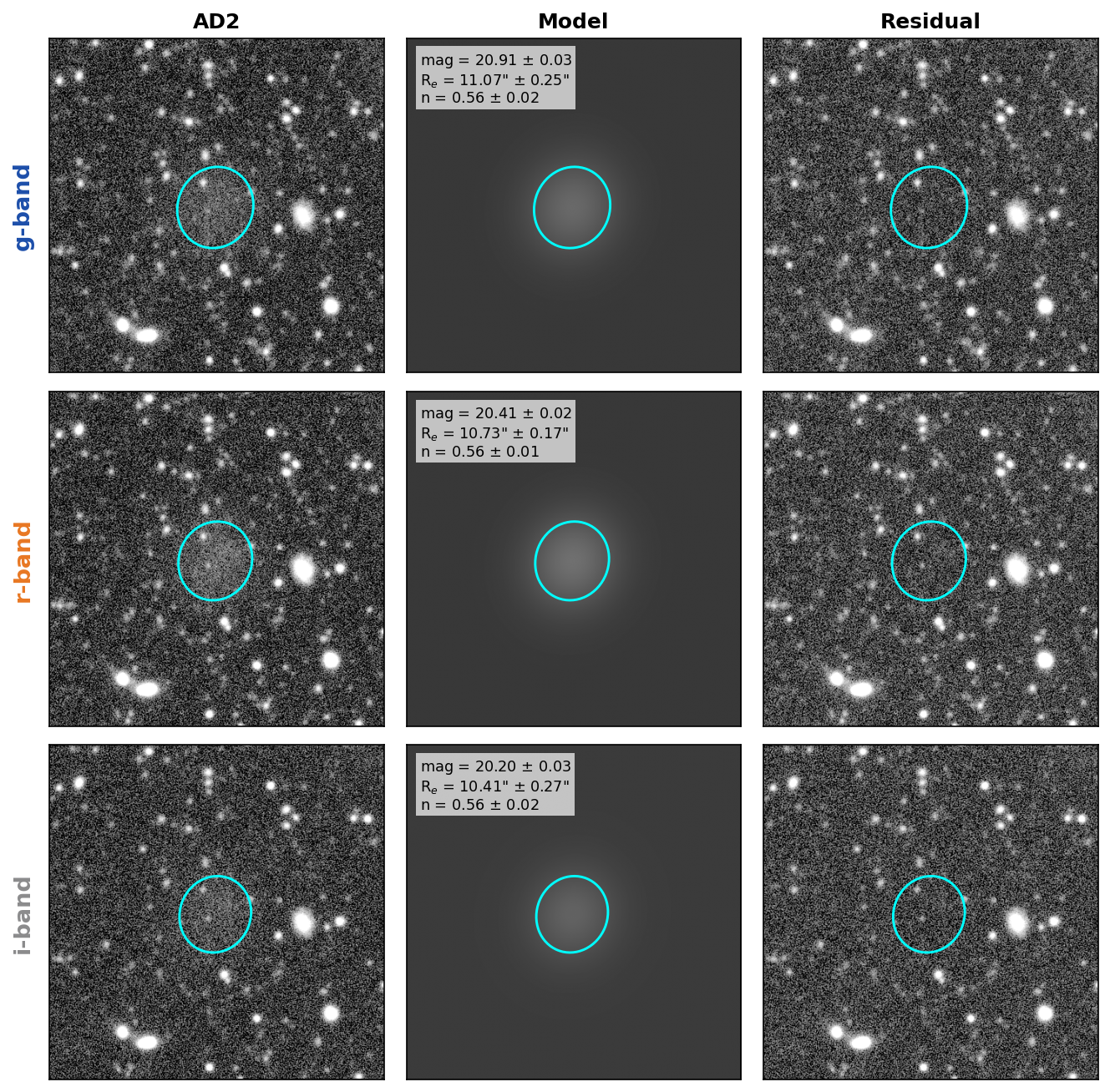}
        \caption{Same as above, for AD~2.}
        \label{fig:AD2_galfitm}
\end{figure*}

\begin{figure*}[ht!]
     \centering
        \includegraphics[width=1\linewidth]{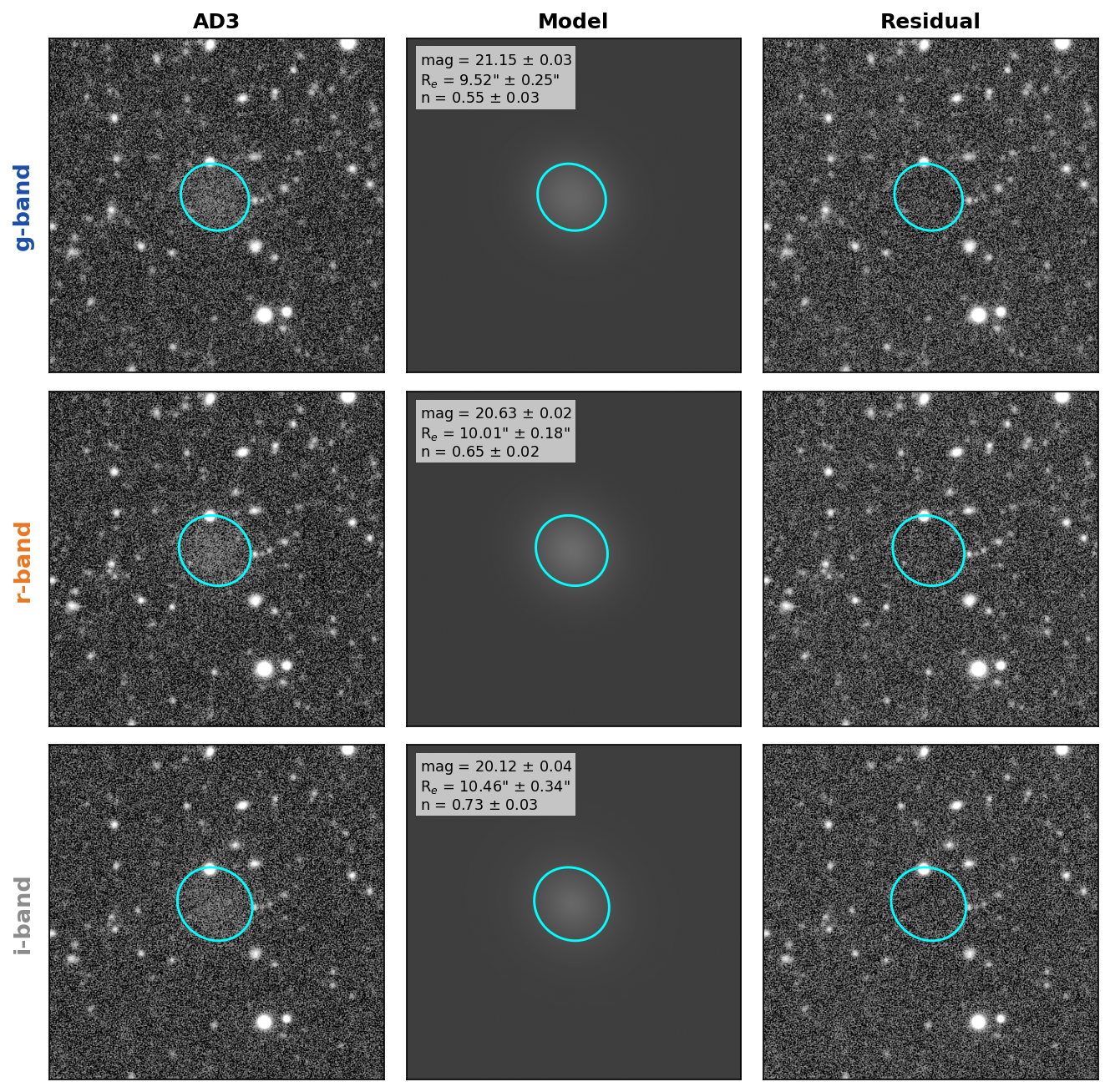}
        \caption{Same as above, for AD~3.}
        \label{fig:AD3_galfitm}
\end{figure*}

\begin{figure*}[ht!]
     \centering
        \includegraphics[width=1\linewidth]{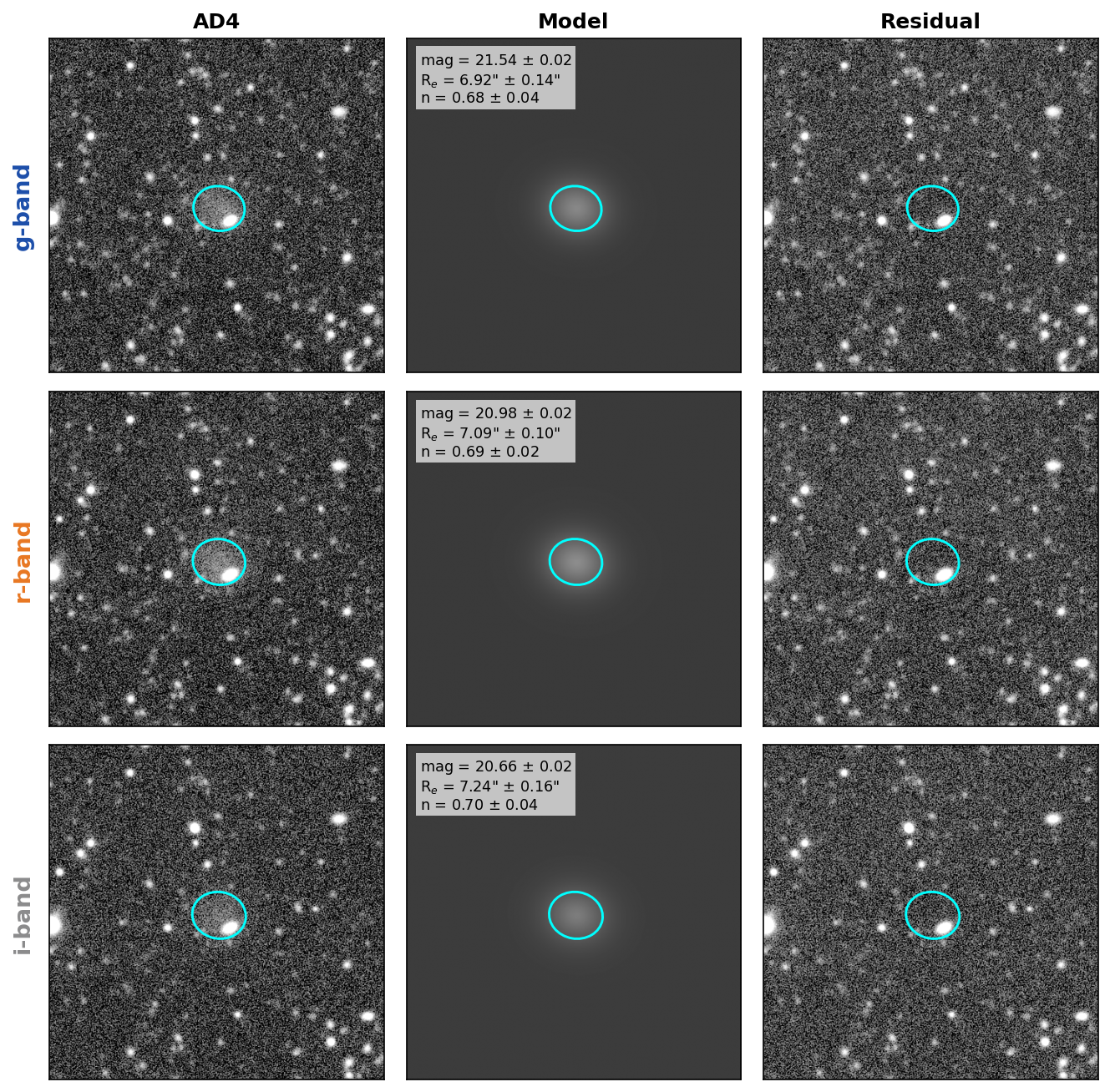}
        \caption{Same as above, for AD~4.}
        \label{fig:AD4_galfitm}
\end{figure*}

\begin{figure*}[ht!]
     \centering
        \includegraphics[width=1\linewidth]{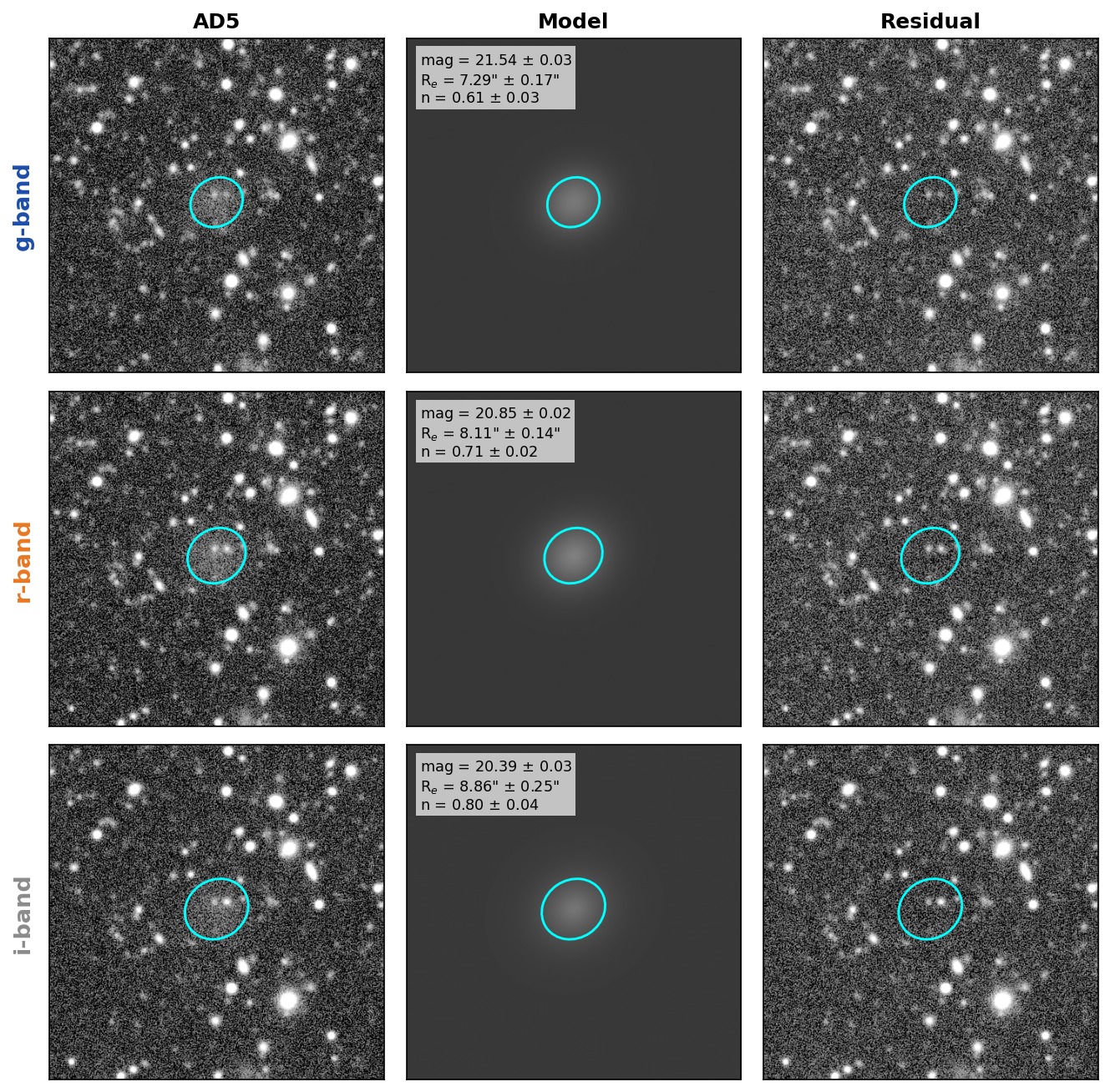}
        \caption{Same as above, for AD~5.}
        \label{fig:AD5_galfitm}
\end{figure*}

\begin{figure*}[ht!]
     \centering
        \includegraphics[width=1\linewidth]{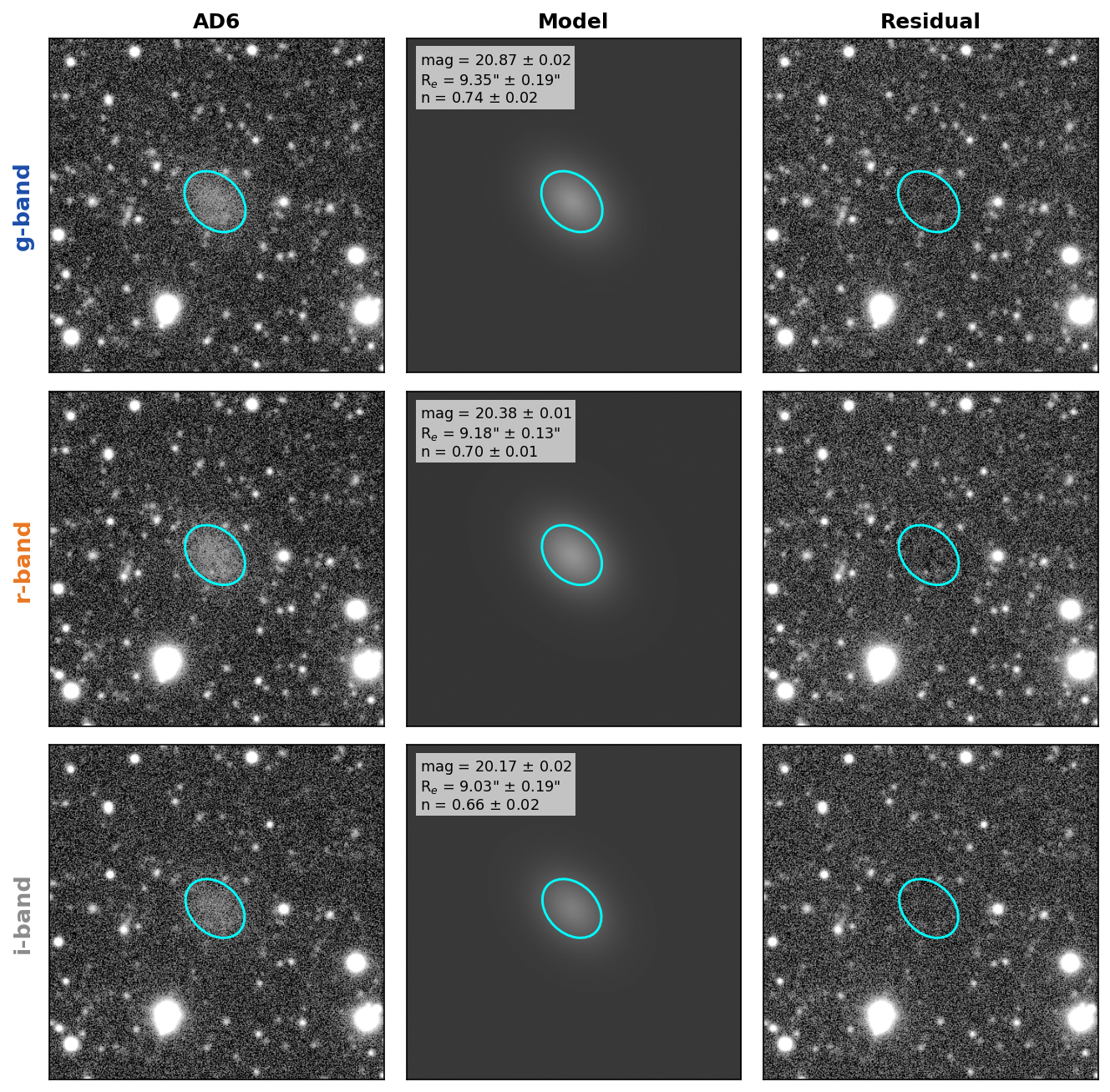}
        \caption{Same as above, for AD~6.}
        \label{fig:AD6_galfitm}
\end{figure*}

\begin{figure*}[ht!]
     \centering
        \includegraphics[width=1\linewidth]{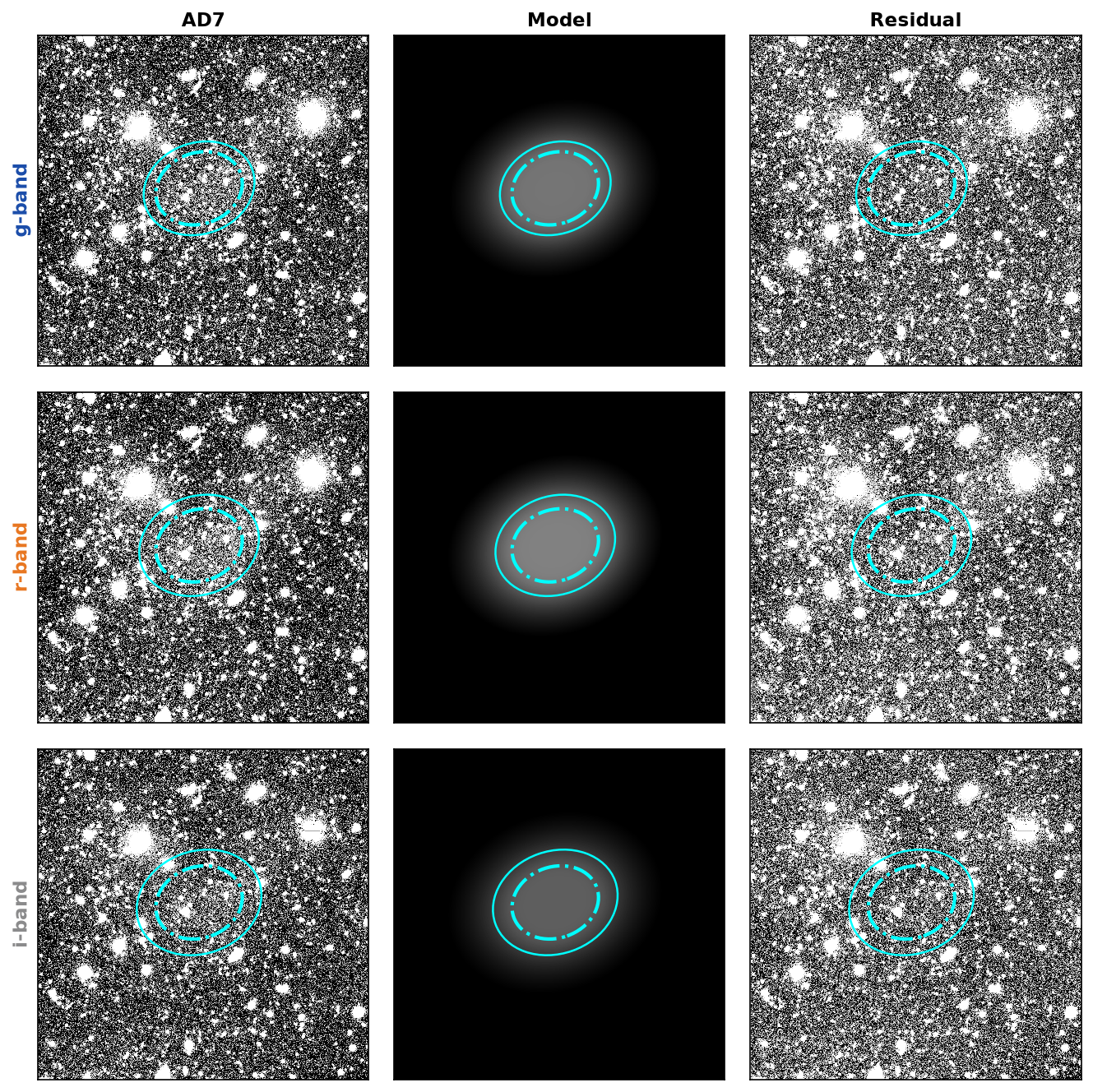}
        \caption{Same as above, but fitting with \texttt{IMFIT} core-Sérsic model for AD~7. The dotted-dashed ellipse shows the break radius $r_\mathrm{b}$, and the solid ellipse shows the effective radius of the core-Sérsic model $r_\mathrm{e, CS}$.}
        \label{fig:AD7_galfitm}
\end{figure*}

\end{appendix}
\end{document}